\documentclass{article}

\usepackage{arxiv}

\usepackage[utf8]{inputenc} 
\usepackage[T1]{fontenc}    
\usepackage{hyperref}       
\usepackage{url}            
\usepackage{booktabs}       
\usepackage{amsfonts}       
\usepackage{nicefrac}       
\usepackage{microtype}      
\usepackage{lipsum}		
\usepackage{graphicx}
\usepackage{natbib}
\usepackage{doi}
\usepackage{bm}
\usepackage{amsmath}
\usepackage{float}

\DeclareMathOperator*{\argmin}{arg\,min}

\title{Bayesian Index Models for Heterogeneous Treatment Effects}

\date{} 					

\author{Hyung Park, Danni Wu, Eva Petkova, Thaddeus Tarpey, R. Todd Ogden}

\author{ \href{https://orcid.org/0000-0002-8994-9583}{\includegraphics[scale=0.06]{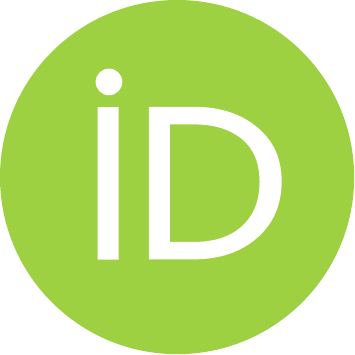}\hspace{1mm}Hyung Park,}{ Danni Wu, Eva Petkova, Thaddeus Tarpey} 
\\
	Department of Population Health\\
	New York University, 
	New York, NY 10016 \\
	\texttt{parkh15@nyu.edu} \\
	\And
	R. Todd Ogden \\
	Department of Biostatistics\\
	Columbia University\\
	New York, NY 10032
}

\renewcommand{\shorttitle}{\textit{arXiv} Bayesian index models for heterogeneous treatment effects}

\hypersetup{
pdftitle={A template for the arxiv style},
pdfsubject={q-bio.NC, q-bio.QM},
pdfauthor={David S.~Hippocampus, Elias D.~Striatum},
pdfkeywords={Bayesian single index models, Heterogeneous treatment effects, Precision medicine},
}

\begin{document}
\maketitle

\begin{abstract} 
The general idea of this article is to 
  develop a Bayesian model with a flexible link function connecting an exponential family treatment response to a linear combination of covariates and a treatment indicator 
  and the interaction between the two. Generalized linear models allowing data-driven link functions are often called "single-index models,” and among popular semi-parametric modeling methods. In this article, we will focus on modeling heterogeneous treatment effects, with the goal of developing a treatment benefit index (TBI) incorporating prior information from historical data. This treatment benefit index can be useful for stratifying patients according to their predicted treatment benefit levels and can be especially useful for precision health applications. 
  The proposed method is applied to a COVID-19 treatment study. 
\end{abstract}


 \section{Introduction}
 In this paper, we develop 
 a Bayesian estimation of single-index models  \citep{Antoniadis,  Choi.2011, Poon2013, Dhara2020} for heterogeneous treatment effects, to optimize individualized treatment rules (ITRs) \citep[e.g.,][]{QianAndMurphy, LU.2011, MC, A.learning.Shi2016, A.learning.Jeng2018, Zhao.2012, Zhao.2015, Song.2015, Laber.Zhao.2015, Laber2018}. 
We consider a treatment variable $A$ taking a value in $\{0, 1\}$ with the associated randomization probabilities $\{\pi_0, \pi_1\}$, 
in the context of randomized clinical trials (RCTs). 
 The observable potential outcomes 
 are $(Y^{(0)},  Y^{(1)})$. 
 Depending on $A$, the observed outcome is 
 $Y = (1-A) Y^{(0)} +  A Y^{(1)}$, 
 with the outcome $Y$  assumed to be a member of the exponential family. 
 Without loss of generality, we assume that a small value of $Y$ is desired. 
 On the population level, this means that a small value of 
 $h(E[Y])$ is desired, where 
 $h(\cdot)$ denotes the canonical link of the assumed exponential family distribution. 
 The covariate $\bm{X} \in \mathbb{R}^p$ are observed pretreatment measurements and predictors of $(Y^{(0)},  Y^{(1)})$. 
Our goal is to utilize the information in $\bm{X}$ 
to 
develop an ITR 
optimizing the value of $h(E[Y])$ for future patients. 

 \section{Method}
 
 \subsection{Optimal individualized treatment rules} \label{ITR}
 In this subsection, we define an optimal ITR. 
The Bayes decision 
$a^\ast: \bm{x} \mapsto \{0,1\}$ 
minimizes, over treatment decision $a \in \{0,1\}$, 
 the posterior expected loss for a patient with baseline measures $\bm{X} = \bm{x}$. 
 Let us define the loss function for making treatment decision $a$ as: 
\begin{equation} \label{value}
L(a, \bm{\theta}, \bm{x}) = h(E[ Y^{(a)} | \bm{\theta}, \bm{x}]),
\end{equation} 
where $\bm{\theta}$ collectively represents the parameters characterizing 
the relationship between the potential treatment outcomes $Y^{(a)}$ and predictors $\bm{X}$. 
In (\ref{value}), 
$E[ Y^{(a)} | \bm{\theta}, \bm{x}]  =  (1-a)  E[ Y^{(0)} | \bm{\theta}, \bm{x} ]  + a  E[ Y^{(1)} | \bm{\theta}, \bm{x} ]$ 
is the expected outcome under treatment assignment $a$. 
Let $\mathcal{D} = \{(Y_i, A_i, \bm{X}_i), \  i=1,\ldots,n \}$ 
collectively denotes the observed data.

Viewing the loss $L(a, \bm{\theta}, \bm{x})$ in (\ref{value}) as a function of $a$ for a patient with pretreatment characteristic  $\bm{x}$,  
 the optimal Bayes decision $a^{\ast}(\bm{x})$ will minimize 
the posterior expected loss given $\bm{x}$, 
i.e., 
$$
a^{\ast}(\bm{x}) \ = \  \argmin_{a \in \{0,1\}}  \ E_{\bm{\theta} | \bm{x}, \mathcal{D}}[ L(a, \bm{\theta}, \bm{x}) ],
$$ 
where the expectation is taken with respect to the posterior distribution of $\bm{\theta}$ (given the observed data $\mathcal{D}$). 
 In particular, if we define the loss contrast 
 $\Delta(\bm{\theta}, \bm{x}) =  L(a=1, \bm{\theta}, \bm{x}) - L(a=0, \bm{\theta}, \bm{x})$, 
then the above optimal Bayes decision $a^{\ast}(\bm{x})$ is equivalently to: 
   \begin{equation} \label{loss.contrast}
   a^{\ast}(\bm{x}) \ =\  \mathbb{I}\{ E_{\bm{\theta} | \bm{x}, \mathcal{D}}[ \Delta(\bm{\theta}, \bm{x})] < 0\}, 
    \end{equation}
    which we define as the optimal ITR.  
We will utilize the following 
 standard causal inference assumptions \citep{Rubin2005}: 1) consistency; 2) no unmeasured confoundedness; 3) positivity, we refer to \cite{Rubin2005} for the details. 
Under those standard assumptions, 
we can write  $\Delta(\bm{\theta}, \bm{x})$ in (\ref{loss.contrast}) as:  
 $\Delta(\bm{\theta}, \bm{x}) = h\{ E[ Y | A= 1, \bm{\theta}, \bm{x} ] \} -  h\{E[ Y | A= 0, \bm{\theta}, \bm{x} ]\}$. 
 Therefore, we can infer the optimal Bayes decision (\ref{loss.contrast})  
based on posterior inference on 
the canonical parameter 
 $h\{E[ Y | A, \bm{\theta}, \bm{X} ]\}$ of the exponential family response $Y$. 
In the following subsection, we will describe 
how we specify the model 
 for 
 the distributions of 
 $(Y | A, \bm{\theta}, \bm{X} )$ and of $\bm{\theta}$, 
 for the estimation of the optimal ITR (\ref{loss.contrast}).  
 


 \subsection{Model  and prior specification} 
 
\subsubsection{Model} 
 Let $\bm{Y} = (Y_1,\ldots, Y_n)^\top$ be a vector of the treatment 
 outcomes, with $Y_i$ following 
 an exponential family distribution with density 
\begin{equation} \label{eq2}
\begin{aligned} 
 f(Y_i | \eta_i, \phi)  
 &= \exp\left\{ \phi^{-1 }[Y_i \eta_i - b(\eta_i)] + c(Y_i, \phi )   \right\}  \\ 
 \eta_i 
 &= 
\bm{X}_i^\top \bm{m} 
+  
  g(\bm{X}_i^\top   \bm{\beta}, A_i), 
\end{aligned} 
\end{equation} 
where the unknown parameters, 
which we collectively denote as $\bm{\theta}$, 
will be estimated in a Bayesian framework.  
In (\ref{eq2}), 
 $b(\cdot)$ and $c(\cdot)$ are known functions  
specific to the given member of the 
exponential family, 
and $\phi > 0$ is an unknown dispersion parameter 
($\phi =1$ specializes to a one-parameter exponential family distribution for the response). 

The canonical parameter $\eta \in \mathbb{R}$ in (\ref{eq2}) represents the location of the assumed exponential family response $Y$, 
which is related to the loss function $L(a, \bm{\theta}, \bm{x}) $ in (\ref{value}) through the equations 
$
 \eta =
 h(E[Y | \bm{\theta}, \bm{x}, a]) = 
 h(E[ Y^{(a)} | \bm{\theta}, \bm{x}]) 
$, 
  under the standard causal inference assumptions. 
 
 The first term $\bm{X}^\top\bm{m}$ in (\ref{eq2}) represents the pre-treatment covariates $\bm{X}$'s ``main'' effect, 
and the second term $g(\bm{X}^\top   \bm{\beta}, A)$ is 
 the $\bm{X}$-by-$A$ interaction effect, 
characterized by an unspecified treatment $a$-specific smooth function $g(u, a)$ $(a=0,1)$ 
which is a function of 
a linear projection $u = \bm{X}^\top \bm{\beta} \in \mathbb{R}$, satisfying $\lVert \bm{\beta} \rVert=1$. 
The projection $\bm{X}^\top \bm{\beta}$  provides a dimension reduction specifically for the $\bm{X}$-by-$A$ interaction effect. 
In (\ref{eq2}), 
 we shall impose 
an identifiability condition 
\begin{equation}\label{constraint0}
E
[ g(\bm{X}^\top \bm{\beta}, A) |\bm{X}] = 0, 
\end{equation} 
which  
separates 
the component   
 $g(\bm{X}^\top\bm{\beta},A)$ 
from the component   
 $\bm{X}^\top \bm{m}$  within $\eta$. 
In (\ref{eq2}), 
the covariates $\bm{X}$  entering into $\bm{X}^\top \bm{m}$ and those into $g(\bm{X}^\top \bm{\beta}, A)$ do not need to be the same.
The model (\ref{eq2}) with the identifiability condition (\ref{constraint0}) is more suitable 
to conduct a posterior inference for heterogeneous treatment effects 
than the model 
$\eta = \bm{X}^\top\bm{m} + g(\bm{X}^\top \bm{\beta}) A$, 
because this particular parametrization (\ref{eq2})  is invariant of the choice of coding of $A$. 
In the latter model,  the choice of the treatment coding can meaningfully impact posterior inferences 
because $\bm{X}^\top \bm{m}$ and $g(\bm{X}^\top\bm{\beta})$ alias one another.  
 On the other hand, if we use the model (\ref{eq2}) with the condition (\ref{constraint0}), 
there is no issue of aliasing of the treatment effects,  
since $g(\bm{X}^\top\bm{\beta}, A)$ is designed to be orthogonal to  $\bm{X}^\top \bm{m}$, even when $\bm{X}^\top \bm{m}$ is misspecified. 
For an individual with baseline characteristics $\bm{x}$, 
the loss contrast $\Delta(\bm{\theta}, \bm{x}) $ in 
(\ref{loss.contrast}) 
under model (\ref{eq2}) is 
\begin{equation} \label{contrast0}
\Delta(\bm{\theta}, \bm{x})  = g(\bm{x}^\top \bm{\beta}, A=1) - g(\bm{x}^\top   \bm{\beta}, A=0), 
\end{equation} 
where 
only the parameters 
$g$ and $\bm{\beta}$ 
(and not $\bm{m}$ and $\phi$)  in (\ref{eq2})  
are necessary for estimating the ITR (\ref{loss.contrast}), 
hence we will focus on the estimation of $g$ and $\bm{\beta}$. 
Given (\ref{loss.contrast}), we can now introduce a 
``treatment benefit index'' (TBI) probability, 
\begin{equation} \label{TBI} 
\mbox{TBI}(\bm{x}) := P(\Delta(\bm{\theta}, \bm{x})   < 0) 
\in [0,1],
\end{equation}  
where the probability 
 is evaluated with respect to the posterior distribution of $\bm{\theta}$. 
The optimal Bayes decision $a^\ast(\bm{x})$ 
in (\ref{loss.contrast}) 
is then $a^\ast(\bm{x}) = \mathbb{I}(\mbox{TBI}(\bm{x}) > 0.5)$. 
Since a large (small) value of  the TBI 
will indicate a large (small) value of relative ``benefit'' from taking the active treatment $A=1$ compared to $A=0$, 
the $\mbox{TBI}$ in (\ref{TBI}) 
 constructs 
 a ``gradient'' of treatment benefit  ranging from $0$ to $1$,  comparing $A=1$ vs $A=0$,
with respect to the covariate value $\bm{x}$. 
Furthermore,  
for each treatment condition $A=a$, 
we can obtain a prediction of the expected outcome  $h^{-1}\{\bm{x}^\top \bm{m} + g(\bm{x}^\top \bm{\beta}, a)\}$
 based on the posterior distribution of the parameters $\bm{\theta}$, 
 for each $\bm{x}$.

 \subsubsection{Representation of the link function $g$} \label{representation.g}

Following \cite{Antoniadis}, we represent the flexible function $g(\cdot, a)$ of (\ref{eq2}) 
 with cubic splines with the $B$-spline basis.   Using $B$-splines   
  is appealing because the basis functions are strictly local, 
 as each basis function is only non-zero over the intervals between 5 adjacent knots \citep{Eilers1996}.   
For each fixed $\bm{\beta} \in \mathbb{R}^p$,  the flexible function $g$ is represented as: 
\begin{equation}\label{f.representation2}
g(\bm{\beta}^\top \bm{x}_i, a_i)  = \tilde{\bm{\psi}}_{\bm{\beta}}(\bm{\beta}^\top \bm{x}_i)^\top \tilde{\bm{\gamma}}_{a_i}    \quad (i=1,\ldots,n)  
\end{equation}
for some
fixed $l$-dimensional basis $\tilde{\bm{\psi}}(\cdot) \in \mathbb{R}^l$ 
(e.g., $B$-spline basis on evenly spaced knots on a bounded range of $\{\bm{\beta}^\top \bm{x}_i \}_{i=1}^n$) 
and a set of unknown treatment $a$-specific basis coefficients 
$\{ \tilde{\bm{\gamma}}_a  \in \mathbb{R}^l \}_{ a \in \{0,1\}}$. 

Given representation (\ref{f.representation2}) for the function $g$ given any $\bm{\beta}$, 
the identifiability constraint $\mathbb{E}[ g(\bm{\beta}^\top \bm{X}, A) |\bm{X}]=0$ 
is implied by the linear constraint 
\begin{equation} \label{constraint}
\pi_0  \tilde{\bm{\gamma}}_0 + \pi_1 \tilde{\bm{\gamma}}_1  
= \bm{\pi} \tilde{\bm{\gamma}} = \bm{0},
 \end{equation} 
where $\bm{\pi} = [\pi_1 \bm{I}_l; \pi_2 \bm{I}_l]$  
is the $l \times 2l$ matrix (in which $\bm{I}_l$ denotes the $l \times l$ identity matrix)  
and 
$\tilde{\bm{\gamma}}  = (\tilde{\bm{\gamma}}_0^\top, \tilde{\bm{\gamma}}_1^\top) \in \mathbb{R}^{2l}$, an unknown vector. 
To represent (\ref{f.representation2}) in matrix notation, 
let the $n \times l$ matrices $\tilde{\bm{D}}_{\bm{\beta},a}$ $(a=0,1)$ denote the evaluation matrices of the basis function 
$\tilde{\bm{\psi}}_{\bm{\beta}}(\cdot)$ on $\{\bm{\beta}^\top \bm{x}_i \}_{i=1}^n$, 
specific to the treatment $A=a$ $(a=0,1)$, 
whose $i$th row is the 
$1 \times l$ vector $\tilde{\bm{\psi}}_{\bm{\beta}}(\bm{\beta}^\top \bm{x}_i )^\top$ 
if $A_i= a$, and a row of zeros $\bm{0}^\top$ if $A_i \ne a$. Then, 
the column-wise concatenation of the design matrices 
$\{ \tilde{\bm{D}}_{\bm{\beta},a} \}_{a\in\{0,1\}}$, i.e., 
the $n \times 2l$ matrix $\tilde{\bm{D}}_{\bm{\beta}} = [\tilde{\bm{D}}_{\bm{\beta},0};  \tilde{\bm{D}}_{\bm{\beta},1}]$, 
defines the model matrix associated with 
$\tilde{\bm{\gamma}} \in \mathbb{R}^{2l}$.  
Then, we can represent the function $g$ in (\ref{f.representation2}) 
based on the sample data, by the length-$n$ vector: 
 $\bm{g} = \tilde{\bm{D}}_{\bm{\beta}} \tilde{\bm{\gamma}}$. 
 
 The linear constraint (\ref{constraint}) on $\tilde{\bm{\gamma}}$ 
 can be conveniently absorbed into the model matrix 
 $\tilde{\bm{D}}_{\bm{\beta}}$ by
reparametrization, as we describe next.  
We can find a $2l \times l$  basis matrix $\bm{Z}$, such that if we set 
$\tilde{\bm{\gamma}} = \bm{Z} \bm{\gamma}$ for any arbitrary vector  
 $\bm{\gamma} \in \mathbb{R}^l$, then the vector $\tilde{\bm{\gamma}} \in \mathbb{R}^{2l}$ automatically satisfies
the constraint (\ref{constraint}). 
 Such a basis matrix $\bm{Z} $  can be constructed by a QR decomposition of the matrix 
 $\bm{\pi}^\top$. 
 Then representation 
  $\bm{g} = \tilde{\bm{D}}_{\bm{\beta}} \tilde{\bm{\gamma}}$ 
  can be  reparametrized, in terms of the unconstrained vector 
  $\bm{\gamma} \in \mathbb{R}^l$, by replacing 
  $\tilde{\bm{D}}_{\bm{\beta}}$ 
 with the reparametrized model matrix 
 $\bm{D}_{\bm{\beta}} = \tilde{\bm{D}}_{\bm{\beta}} \bm{Z}$, yielding the representation 
 $\bm{g} = \bm{D}_{\bm{\beta}} \bm{\gamma}$. 
 
 Once we perform inference on $\bm{\gamma}$, 
 we can also consider inference on the transformed parameter 
 $\tilde{\bm{\gamma}} = \bm{Z} \bm{\gamma}$, 
 from which we can make inference on the functions $g(\bm{\beta}^\top \cdot, a) =  \tilde{\bm{\psi}}_{\bm{\beta}}(\bm{\beta}^\top \cdot)^\top \tilde{\bm{\gamma}}_{a}$ $(a=0,1)$.

\subsubsection{Prior specification} \label{prior}

How we specify priors for $\bm{\beta}$, $\bm{m}$ and $\bm{\gamma}$ 
is given in this subsection.



\begin{itemize}
\item
For the distribution of 
 $\bm{\beta}$,  
we will use 
von Mises-Fisher with concentration parameter $\lambda_{prior} >0$ and direction parameter $\bm{\beta}_0 \in \mathbb{R}^p$ with $\lVert \bm{\beta}_0 \rVert=1$,  
\begin{equation} \label{priortheta}
P(\bm{\beta}) \propto \exp(\lambda_{prior} \bm{\beta}^\top \bm{\beta}_0), 
 \end{equation}
 a probability distribution for $\bm{\beta}$ on the $(p-1)$-unit-sphere in $\mathbb{R}^p$. 

\item 
We will use 
$\bm{m} \sim N(\bm{m}_0,  \bm{Q})$, 
for some vector $\bm{m}_0 \in \mathbb{R}^p$ and $p \times p$ positive definite matrix $\bm{Q}$.  
 

%

\item 
Since the domain of the function $g$ in (\ref{eq2}) depends on $\bm{\beta}$, 
the prior on $g$ will depend on $\bm{\beta}$.  
For each fixed $\bm{\beta}$, 
we 
will use data-dependent empirical Bayes  prior for $\bm{\gamma} \in \mathbb{R}^l$ 
\begin{equation} \label{priorbeta}
\bm{\gamma} \sim N(\widehat{\bm{\gamma}}_{\bm{\beta}},  \   (\bm{D}_{\bm{\beta}}^{\top} \bm{W} \bm{D}_{\bm{\beta}})^{-1}  ),
 \end{equation}
where the mean $   \widehat{\bm{\gamma}}_{\bm{\beta}}$ is 
\begin{equation} \label{pls}
   \widehat{\bm{\gamma}}_{\bm{\beta}}
  =
 (\bm{D}_{\bm{\beta}}^{\top} \bm{W} \bm{D}_{\bm{\beta}} + \rho\bm{I})^{-1}  \bm{D}_{\bm{\beta}}^{\top} \bm{W} 
  \bm{Z}. 
 \end{equation} 
How we specify the tuning parameter $\rho >0$ 
 and  
 $\bm{Z}$ and $\bm{W}$ in  
 (\ref{pls}) 
is given in the next subsection. 
An advantage of using the prior (\ref{priorbeta})  
is that 
it allows us to analytically integrate $\bm{\gamma}$ out of the 
joint posterior $P(\bm{\beta}, \bm{\gamma} | \bm{m}, \bm{Y})$,  
facilitating 
the Gibbs sampling of $\bm{\beta}$. 
   For  simplicity of the notation in the next subsection, 
   let us write 
  \begin{equation*}\label{thetarho}
   \bm{\Sigma}_\rho := (\bm{D}_{\bm{\beta}}^\top \bm{W} \bm{D}_{\bm{\beta}} + \rho \bm{I})^{-1},
   \end{equation*} 
   and 
   $\bm{\Sigma}_0 = (\bm{D}_{\bm{\beta}}^\top \bm{W}  \bm{D}_{\bm{\beta}})^{-1}$, 
   which is a special case of $\bm{\Sigma}_\rho$ at $\rho =0$. 
   Then the prior 
    in (\ref{priorbeta}) is simply written as 
    \begin{equation}\label{priorbeta2}
   \bm{\gamma} \sim N( \bm{\Sigma}_\rho \bm{D}_{\bm{\beta}}^{\top} \bm{W} \bm{Z},
    \    \bm{\Sigma}_0).    
   \end{equation}


%
%

\end{itemize}



\subsection{Posterior computation} 
  
To conduct posterior inference on $(\bm{m}, \bm{\beta}, \bm{\gamma})$, 
we will simulate samples from the joint posterior $P(\bm{m}, \bm{\beta}, \bm{\gamma} | \bm{Y})$.
Since 
 it is difficult to draw samples directly from this joint posterior, 
we will use a Metropolis-Within-Gibbs algorithm. 
The Gibbs algorithm will iterate between the following two Steps: 
\textit{Step 1})  sample $\bm{m}$ from $P(\bm{m} | \bm{\beta}, \bm{\gamma}, \bm{Y})$; and 
\textit{Step 2})  sample $(\bm{\beta}, \bm{\gamma})$ from $P(\bm{\beta}, \bm{\gamma}  | \bm{m}, \bm{Y})$. 
In \textit{Step 2}, since the joint conditional $P(\bm{\beta}, \bm{\gamma}  | \bm{m}, \bm{Y})$ does not have a convenient form, 
we will employ a Metropolis-Hastings step. 
 

  
 \subsubsection{Conditional posteriors}
  
\begin{enumerate}
\item
Derivation of $(\bm{m} |\bm{\beta}, \bm{\gamma},\bm{Y})$. 
For fixed $\bm{\beta}$ and  $\bm{\gamma}$, we will approximate the conditional distribution of  $(\bm{m} |\bm{\beta}, \bm{\gamma},\bm{Y})$.  
Specifically, we will quadratically approximate the log likelihood of $\bm{m}$, centered at its mode $\check{\bm{m}}$. 
 To find the mode  $\check{\bm{m}}$, we 
will use a Fisher scoring, 
  iteratively updating the center of the quadratic approximation. 
 Given $\bm{\beta}$ and $\bm{\gamma}$,  at the convergence of the Fisher scoring, we will have 
the \textit{adjusted response} vector $\check{\bm{Z}} = (\check{z}_1,\ldots, \check{z}_n)^\top \in \mathbb{R}^n$ 
  where $\check{z}_i := h^\prime(\check{\mu}_i)(y_i- \check{\mu}_i) + \check{\eta}_i$,  
in which $\check{\eta}_i = \check{\bm{m}}^\top \bm{x}_i +  \bm{\psi}(\bm{\beta}^\top \bm{x}_i)^\top \bm{\gamma}$ and 
  $\check{\mu}_i = h^{-1}(\check{\eta}_i)$, 
   and we will have the $n \times n$ \textit{weight} matrix 
   $\check{\bm{W}} = \mbox{diag}(\check{w}_i)$.  
   As a result, for fixed $\bm{\beta}$ and  $\bm{\gamma}$,  
   the negative log likelihood of $\bm{m}$ is approximately represented in terms of a weighted least squares (WLS) objective function (up to a constant of proportionality): 
  \begin{equation}\label{wls2}
  \begin{aligned}
\sum_{i=1}^n \check{w}_i (\check{z}_i - \bm{m}^\top \bm{x}_i)^2 
&=  (\check{\bm{Z}}  -  \bm{X} \bm{m}  )^{\top} \check{\bm{W}} (\check{\bm{Z}} -  \bm{X} \bm{m}) \\
&= \mbox{SSE} + (\bm{m} - (\bm{X}^\top \check{\bm{W}} \bm{X})^{-1} \bm{X}^\top \check{\bm{W}} \check{\bm{Z}})^\top  
\bm{X}^\top \check{\bm{W}}  \bm{X} 
(\bm{m} - (\bm{X}^\top \check{\bm{W}} \bm{X})^{-1} \bm{X}^\top \check{\bm{W}} \check{\bm{Z}}), 
\end{aligned}
\end{equation} 
where the term  
$\mbox{SSE} := (\check{\bm{Z}} - \bm{X} (\bm{X}^\top \check{\bm{W}} \bm{X})^{-1} \bm{X}^\top \check{\bm{W}} \check{\bm{Z}} )^\top \check{\bm{W}} (\check{\bm{Z}} - \bm{X} (\bm{X}^\top \check{\bm{W}} \bm{X})^{-1} \bm{X}^\top \check{\bm{W}} \check{\bm{Z}})$ 
does not involve $\bm{m}$. 
Given the prior $\bm{m} \sim N(\bm{m}_0, \bm{Q})$ 
and the second term in the WLS objective (\ref{wls2}),  
 the compionents associated with 
$\bm{m}$ in the negative log posterior are (up to a constant of proportionality):
$$
(\bm{m} - \bm{m}_0)^\top \bm{Q}^{-1} (\bm{m} - \bm{m}_0) +  
(\bm{m} - (\bm{X}^\top \check{\bm{W}} \bm{X})^{-1} \bm{X}^\top \check{\bm{W}} \check{\bm{Z}})^\top  
\bm{X}^\top \check{\bm{W}}  \bm{X} 
(\bm{m} - (\bm{X}^\top  \check{\bm{W}} \bm{X})^{-1} \bm{X}^\top \check{\bm{W}} \check{\bm{Z}}). 
$$
This 
indicates that 
the conditional posterior  for $\bm{m}$ is:  
\begin{equation} \label{conditional.m}
P(\bm{m}| \bm{\beta}, \bm{\gamma}, \bm{Y}) \ = \
N((\bm{Q}^{-1} + \bm{X}^\top \check{\bm{W}} \bm{X})^{-1} 
(\bm{Q}^{-1} \bm{m}_0 + 
 \bm{X}^\top \check{\bm{W}} \check{\bm{Z}} ), 
 \ 
 (\bm{Q}^{-1} + \bm{X}^\top \check{\bm{W}} \bm{X})^{-1}).  
\end{equation} 



 \item
 Derivation of $(\bm{\beta}, \bm{\gamma} | \bm{m}, \bm{Y})$.  
Given that the joint conditional $P(\bm{\beta}, \bm{\gamma} | \bm{m}, \bm{Y})
 = P(\bm{\beta} | \bm{m}, \bm{Y})P(\bm{\gamma} | \bm{\beta}, \bm{m}, \bm{Y})$, 
 we will first sample $\bm{\beta}$ from $P(\bm{\beta} | \bm{m}, \bm{Y})$ and then $\bm{\gamma}$ from $P(\bm{\gamma} | \bm{\beta}, \bm{m}, \bm{Y})$. 
Specifically, following 
\cite{Antoniadis},  
we will use a Metropolis-Hastings algorithm 
to sample $\bm{\beta}$ from
  $p(\bm{\beta} | \bm{m}, \bm{Y})$. 
       However,   this approach employed in \cite{Antoniadis}  cannot be directly applied to our settings, 
            since the outcome $Y_i$ is generally not Gaussian. 
 Thus, we will perform a quadratic approximation 
 of the negative log likelihood of $\bm{\gamma}$, 
 at its mode, which we denote as $\hat{\bm{\gamma}}$, i.e., 
  approximating the likelihood by a normal density in $\bm{\gamma}$ centered at $\hat{\bm{\gamma}}$.
 To find $\hat{\bm{\gamma}}$, we will again conduct a Fisher scoring. 
 For each fixed $\bm{\beta}$ and $\bm{m}$, 
this quadratic approximation 
at the convergence of the Fisher scoring is 
 summarized in the form of 
the WLS objective function (up to a constant of proportionality),  
\begin{equation}\label{wls}
\sum_{i=1}^n w_i (z_i - \bm{\psi}(\bm{\beta}^\top \bm{x}_i)^\top \bm{\gamma}   )^2 
=  (\bm{Z} -  \bm{D}_{\bm{\beta}}   \bm{\gamma}  )^{\top}  \bm{W} (\bm{Z} -  \bm{D}_{\bm{\beta}}   \bm{\gamma}  ), 
\end{equation} 
as a function of $\bm{\gamma}$, 
in which 
$\bm{Z} = (z_1,\ldots,z_n)^\top \in \mathbb{R}^n$ is 
the 
  \textit{adjusted response} vector with $z_i= h^\prime(\hat{\mu}_i)(y_i- \hat{\mu}_i) + \hat{\eta}_i$ 
obtained at the convergence, 
  where 
  $\hat{\eta}_i = \bm{m}^\top \bm{x}_i +  \bm{\psi}(\bm{\beta}^\top \bm{x}_i)^\top\hat{\bm{\gamma}}$ 
  and $\hat{\mu}_i = h^{-1}(\hat{\eta}_i)$, 
   and    $\bm{W} = \mbox{diag}(w_i)$  is the $n \times n$ \textit{weight} matrix 
with  $w_i = 1/\{  (h^\prime(\hat{\mu}_i))^2 V(\hat{\mu}_i)\}$. 
Given the quadratic approximation (\ref{wls}), 
we can write the joint conditional $(\bm{\beta}, \bm{\gamma} | \bm{m},\bm{Y})$:
  \begin{equation} \label{tmp1}
  \begin{aligned}
  P(\bm{\beta}, \bm{\gamma} | \bm{m},\bm{Y}) 
   =   & \ 
P(\bm{Y} | \bm{\gamma}, \bm{\beta}, \bm{m})  P( \bm{\gamma} |  \bm{\beta}, \bm{m}) 
P(\bm{\beta}) 
         \\  
  \propto & \  
 \exp\{ - \frac{1}{2} (\bm{Z} -  \bm{D}_{\bm{\beta}}   \bm{\gamma}  )^{\top}  \bm{W} (\bm{Z} -  \bm{D}_{\bm{\beta}}   \bm{\gamma})\} 
\  P( \bm{\gamma} |  \bm{\beta}, \bm{m}) \ 
     \exp(\lambda_{prior} \bm{\beta}^\top \bm{\beta}_0). 
\end{aligned}
\end{equation}
We will now integrate $\bm{\gamma}$ out of (\ref{tmp1}) to obtain an expression for  $P(\bm{\beta} | \bm{m},\bm{Y})$. 
Utilizing the empirical Bayes prior $P( \bm{\gamma} |  \bm{\beta}, \bm{m})$ specified in  
 (\ref{priorbeta2}), 
we can write the terms involving $\bm{\gamma}$  in 
 (\ref{tmp1}) as: 
  \begin{equation} \label{eq3}
  \begin{aligned}
\propto 
& \exp
\left(
- \frac{1}{2} 
\left\{ 
(\bm{Z} - \bm{D}_{\bm{\beta}}\bm{\gamma})^\top \bm{W} (\bm{Z}- \bm{D}_{\bm{\beta}}\bm{\gamma}) 
+
(\bm{\gamma} - \bm{\Sigma}_\rho \bm{D}_{\bm{\beta}}^{\top} \bm{W} \bm{Z} )^\top \bm{\Sigma}_0^{-1} (\bm{\gamma} - \bm{\Sigma}_\rho \bm{D}_{\bm{\beta}}^{\top} \bm{W} \bm{Z}) 
\right\}
 \right) \\
  \propto 
& 
\exp
\left(
- \frac{1}{2} 
\left\{ 
(\bm{\gamma} - \bm{\Sigma}_0 \bm{D}_{\bm{\beta}}^{\top} \bm{W} \bm{Z}) \bm{\Sigma}_0^{-1} (\bm{\gamma} - \bm{\Sigma}_0 \bm{D}_{\bm{\beta}}^{\top} \bm{W} \bm{Z}) 
+ (\bm{\gamma} - \bm{\Sigma}_\rho \bm{D}_{\bm{\beta}}^{\top} \bm{W} \bm{Z} )^\top \bm{\Sigma}_0^{-1} (\bm{\gamma} - \bm{\Sigma}_\rho \bm{D}_{\bm{\beta}}^{\top} \bm{W} \bm{Z})  
\right\} 
\right)
  \\
  = &
\exp
\left(
- \frac{1}{2} 
\left\{ 
 2 \bm{\gamma}^\top \bm{\Sigma}_0^{-1} \bm{\gamma} - 2 \bm{\gamma}^\top (\bm{I} + \bm{\Sigma}_0^{-1} \bm{\Sigma}_\rho )\bm{D}_{\bm{\beta}}^\top \bm{W} \bm{Z} 
 + \bm{Z}^\top \bm{W} \bm{Z} +  \bm{Z}^\top \bm{W}^\top \bm{D}_{\bm{\beta}} \bm{\Sigma}_\rho \bm{\Sigma}_0^{-1} \bm{\Sigma}_\rho \bm{D}_{\bm{\beta}}^\top \bm{W}\bm{Z}
 \right\}
 \right)  \\
  = &
\exp
\left(
- \frac{1}{2} 
\left\{ 
 2 \bm{\gamma}^\top \bm{\Sigma}_0^{-1} \bm{\gamma} - 2 \bm{\gamma}^\top (\bm{I} + \bm{\Sigma}_0^{-1} \bm{\Sigma}_\rho )\bm{D}_{\bm{\beta}}^\top \bm{W} \bm{Z} 
 + S_1(\bm{\beta})  
  \right\}
 \right),   
\end{aligned}
\end{equation}
where 
$$S_1(\bm{\beta})   = \bm{Z}^\top \bm{W} \bm{Z} +  \bm{Z}^\top \bm{W}^\top \bm{D}_{\bm{\beta}} \bm{\Sigma}_\rho \bm{\Sigma}_0^{-1} \bm{\Sigma}_\rho \bm{D}_{\bm{\beta}}^\top \bm{W}\bm{Z}.$$ 



Specifically, 
using the expression in the third line of (\ref{eq3}), 
 we can analytically integrate $\bm{\gamma}$ out of (\ref{tmp1}), yielding 
  \begin{equation} \label{eq4}
  \begin{aligned}
    P(\bm{\beta} | \bm{m}, \bm{Y}) 
    &= 
     \int  P(\bm{\beta}, \bm{\gamma} |  \bm{m}, \bm{Y}) d\bm{\gamma} \\
   &\propto  \int  
   \exp
\left(
- \frac{1}{2} 
\left\{ 
 2 \bm{\gamma}^\top \bm{\Sigma}_0^{-1} \bm{\gamma} - 2 \bm{\gamma}^\top (\bm{I} + \bm{\Sigma}_0^{-1} \bm{\Sigma}_\rho )\bm{D}_{\bm{\beta}}^\top \bm{W} \bm{Z} 
 + S_1(\bm{\beta})  
  \right\}
 \right) 
 \frac{1}{|\bm{\Sigma}_0|^{1/2}} 
 \exp(\lambda_{prior} \bm{\beta}^\top \bm{\beta}_0) 
   d\bm{\gamma}  \\ 
   &\propto 
\mathcal{L} 
\left[
N(\bm{0}, \bm{\Sigma}_0/2)
\right]\left[- (\bm{I} + \bm{\Sigma}_0^{-1} \bm{\Sigma}_\rho )\bm{D}_{\bm{\beta}} \bm{W} \bm{Z} \right]
     \exp
\left\{
- \frac{1}{2 } S_1(\bm{\beta})  
\right\}  
\exp(\lambda_{prior} \bm{\beta}^\top \bm{\beta}_0) 
\end{aligned}
\end{equation}
where 
$\mathcal{L} 
\left[
N(\bm{0}, \bm{\Sigma}_0/2)
\right]\left[- (\bm{I} + \bm{\Sigma}_0^{-1} \bm{\Sigma}_\rho )\bm{D}_{\bm{\beta}} \bm{W} \bm{Z} \right]$ 
is the  Laplace transform of the density function of $N(\bm{0}, \bm{\Sigma}_0/2)$, 
evaluated at  the parameter 
 $- (\bm{I} + \bm{\Sigma}_0^{-1} \bm{\Sigma}_\rho )\bm{D}_{\bm{\beta}} \bm{W} \bm{Z}$. 
The familiar closed-form expression 
of the 
Laplace transform of Gaussian allows us to write the last line of (\ref{eq4}) as: 
\begin{equation} \label{posterior.theta} 
P(\bm{\beta} | \bm{m},\bm{Y}) \propto 
\exp
\left(
 \frac{1}{4}  \bm{Z}^\top\bm{W}^\top  \bm{D}_{\bm{\beta}}^\top \bm{\Lambda}  \bm{D}_{\bm{\beta}} \bm{W} \bm{Z}
\right) 
\exp
\left\{
- \frac{1}{2} S_1(\bm{\beta})  
\right\} 
\exp(\lambda_{prior} \bm{\beta}^\top \bm{\beta}_0),
\end{equation} 
where 
$\bm{\Lambda} = (\bm{I} + \bm{\Sigma}_\rho \bm{\Sigma}_0^{-1})  \bm{\Sigma}_0  (\bm{I} + \bm{\Sigma}_\rho \bm{\Sigma}_0^{-1})$.  
The expression  (\ref{posterior.theta}) provides a closed form   for 
the approximated $P(\bm{\beta} | \bm{m},\bm{Y})$ 
up to a constant of proportionality, which we will use to conduct 
a random walk Metropolis Markov chain Monte Carlo (MCMC) algorithm. 
The MCMC algorithm to sample $(\bm{\beta} | \bm{m}, \bm{Y})$ is described in the next subsection.  
Given each $\bm{m}$ and a sample $\bm{\beta}$ from $p(\bm{\beta} | \bm{m}, \bm{Y})$, we can sample $\bm{\gamma}$ from $P(\bm{\gamma} |\bm{\beta}, \bm{m}, \bm{Y})$, 
specified by the normal density: 
 \begin{equation} \label{conditional.gamma}
 P(\bm{\gamma} |\bm{\beta}, \bm{m}, \bm{Y}) = N(\frac{\bm{\Sigma}_0}{2}(\bm{I} + \bm{\Sigma}_0^{-1} \bm{\Sigma}_\rho )\bm{D}_{\bm{\beta}}^\top \bm{W} \bm{Z}, \ \frac{\bm{\Sigma}_0}{2} ).
 \end{equation} 
 
 \end{enumerate} 
 



\subsubsection{MCMC algorithm for the posterior sampling} 

In this subsection, we provide the detailed sampling scheme based on the conditional posterior derived in the previous subsection. 

First, we will initialize the model parameters $(\bm{m}, \bm{\beta}, \bm{\gamma})$ by maximum likelihood estimates, that 
maximize the likelihood 
of model (\ref{eq2}) with representation (\ref{f.representation2}) for $g$, where the 
basis coefficient $\bm{\gamma}$ is determined in the form (\ref{pls}), 
with the tuning parameter $\rho > 0$ determined based on the GCV. 
We will then cycle through the following steps. 

\begin{enumerate} 
\item
Sample $\bm{m}$ from $P(\bm{m} | \bm{\beta}, \bm{\gamma}, \bm{Y} )$ in (\ref{conditional.m}) 
given $(\bm{\beta}, \bm{\gamma})$. 

\item
Sample $\bm{\beta}$ from 
 $P(\bm{\beta} | \bm{m}, \bm{Y})$ in (\ref{posterior.theta}) given $\bm{m}$, using the Metropolis algorithm.  
Specifically, given the current state $\bm{\beta}^{\mbox{cur}}$ for $\bm{\beta}$ of the chain, 
a new value $\bm{\beta}^{\mbox{new}}$ is accepted with 
 the acceptance probability  $\min\{1, r \}$, 
where the Metropolis ratio $r$ is given by: 
$$r = \frac{P(\bm{\beta}^{\mbox{new}} | \bm{m}, \bm{Y})} {P(\bm{\beta}^{\mbox{cur}}| \bm{m}, \bm{Y})}, $$ 
using 
the conditional posterior 
 (\ref{posterior.theta}) 
 given $\bm{m}$. 
Some more details on this Metropolis procedure. 

\begin{itemize}

\item
The proposal distribution for $\bm{\beta}^{\mbox{new}}$ was taken to be von Mises-Fisher
 with concentration parameter $\lambda_{prop}> 0$  
 and direction parameter given by the current value $\bm{\beta}^{\mbox{cur}}$. 
 In the simulation example in the next section, we used 
 $\lambda_{prop} = 300$, which gave the acceptance probability of around 0.3 for the proposal, 
 and the sampler 
 appeared to explore the state space for $\bm{\beta}$ adequately. 
We used the R package 
 \texttt{movMF} 
 to generate random samples for $\bm{\beta}$ from von Mises-Fisher distributions. 

 \item
 For the prior distribution of $\bm{\beta}$ in (\ref{priortheta}), 
we can choose $\lambda_{prior} > 0$ (typically in the range of $100 < \lambda_{prior} < 700$), 
depending on the degree of confidence in the prior direction $\bm{\beta}_0$. 

\item
$\rho > 0$ in (\ref{pls})  is another unknown  that controls
 the smoothness of the data-driven function $g$, 
which is crucial to avoid overfitting $g$. 
This will be selected via an empirical Bayes procedure. 
Although, technically, an optimal $\rho$ needs to be selected at each MCMC update, 
in this article, 
we used the generalized cross-validation (GCV) criterion to select $\rho$ 
only at the start of the MCMC run with the frequentist's estimate in place of $\bm{\beta}$, 
to reduce the computational demand and 
since $\rho$ has relatively little effect on the estimation of $\bm{\beta}$.  
 

\end{itemize}



\item

Sample $\bm{\gamma}$ from $P(\bm{\gamma} |\bm{\beta}, \bm{m}, \bm{Y})$ in (\ref{conditional.gamma}) 
given $(\bm{\beta}, \bm{m})$.

\end{enumerate}
  

     
  To obtain the estimated fit 
  $\hat{y}^{\mbox{new}}$
   given a new 
  $\bm{x}^{\mbox{new}}$ and treatment condition $a \in \{0,1\}$, 
 we take the posterior mean of the expected response 
 $h^{-1}\left( \eta \right) = h^{-1}\left( \bm{m}^\top \bm{x}^{\mbox{new}} + \tilde{\bm{\psi}}_{\bm{\beta}}(\bm{\beta}^\top \bm{x}^{\mbox{new}}) \tilde{\bm{\gamma}}_{a} \right)$, 
 based on the posterior sampler output. 
 In particular, we make 
 a treatment decision using 
 the posterior distribution of 
 $\tilde{\bm{\psi}}_{\bm{\beta}}(\bm{\beta}^\top \bm{x}^{\mbox{new}}) (\tilde{\bm{\gamma}}_{1} - \tilde{\bm{\gamma}}_{0})$. 
 Specifically, we will use 
the probability 
$P(\{\tilde{\bm{\psi}}_{\bm{\beta}}(\bm{\beta}^\top \bm{x}^{\mbox{new}}) (\tilde{\bm{\gamma}}_{1} - \tilde{\bm{\gamma}}_{0})< 0\})$ 
as the $\mbox{TBI}(\bm{x}^{\mbox{new}})$,  
which we will utilize to obtain a decision rule 
 $a^{\ast}(\bm{x}^{\mbox{new}}) = \mathbb{I}(\mbox{TBI}(\bm{x}^{\mbox{new}})  > 0.5)$, 
 using the probability threshold of $0.5$. 

\section{Application}

Here we illustrate an application of the proposed model to real data. Specifically, we apply the proposed model to a COVID-19 convalescent plasma (CCP) study
\citep{TroxelEtAl2022}, 
a meta-analysis of pooled individual patient data from 8 randomized clinical trials. 
The goal of this study was to guide CCP treatment recommendations by providing an estimate of a differential treatment outcome when a patient is treated with CCP vs without CCP  \citep{ParkEtAlTBI2022}. A larger differential in favor of CCP would indicate a more compelling reason for recommending CCP. In this context, we aim to discover profiles of patients with COVID-19 associated with different benefit from CCP treatment and use these to optimize treatment decisions.

The study included 2369 hospitalized adults, not receiving mechanical ventilation at randomization, enrolled April 2020 to March 2021. We took complete cases for the analysis. A total of 2287 patients were included, with a mean (SD) age of 60.3 (15.2) years and 815 (35.6\%) women. One of the primary outcomes of the study was the binary variable indicating mechanical ventilation or death (hence $Y = 1$ indicates a bad outcome) at day 14 post-treatment. The patients were randomized to be treated with either CCP $(A=1)$ or control $(A=0)$, i.e., standard of care. 
Pretreatment patient characteristics were collected at baseline.  
In our application, the baseline variables that were used to model the covariates ``main'' effect, i.e.,  
the component associated with the coefficient $\bm{m}$ in model (\ref{eq2})) were age, sex, baseline symptom conditions, age-by-baseline symptom conditions interaction, blood type, the indicators for history of diabetes, pulmonary and cardiovascular disease, 
and days since the symptoms onset. We also included the RCT-specific intercepts and the patients' enrollment quarters as part of the covariates ``main'' effect component. 

Since our goal in this analysis is to investigate the differential treatment effect explained by the baseline variables $\bm{X}$, 
 we will focus on reporting the estimation results of the heterogeneous treatment effect (HTE) term $g(\bm{X}^\top\bm{\beta}, A)$ in model (\ref{eq2})
  and the corresponding treatment effect contrast $\Delta(\bm{x}, \bm{\theta})$ in  (\ref{contrast0}). 
 The patient characteristics $\bm{X}$ included in the HTE term 
 are given in the first column of Table~\ref{tab:table}. 
 The posterior mean of the index coefficients $\bm{\beta} = (\beta_1,\ldots, \beta_7)$,  
 along with the corresponding 
 $95\%$ posterior credible intervals (CrI), 
 are provided in the second column of Table~\ref{tab:table}. 
By examining the posterior CrI, 
  the patient's symptoms severity at baseline, blood type, a history of cardiovascular disease and a history of diabetes 
  appear to be important predictors of HTE.

\begin{figure}[H] 
\begin{center}
\begin{tabular}{c} 
\includegraphics[width=5.3 in, height = 2.5in]{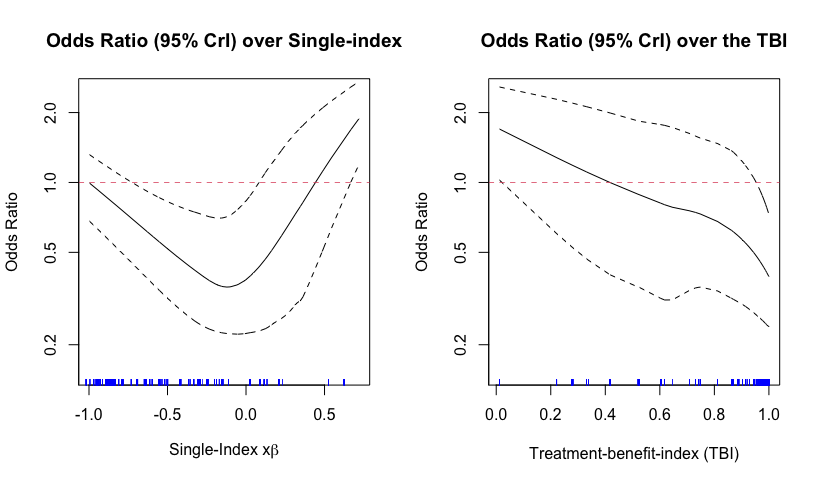}  
\end{tabular}  
\end{center}
  \vspace{-0.2 in}
\caption{
 The left panel   displays the exponentiated version 
 of the estimated individualized treatment effect, 
 the posterior mean of 
 $\Delta(\bm{x}) = g(\bm{x}^\top\bm{\beta}, A=1) - g(\bm{x}^\top\bm{\beta}, A=0)$ in (\ref{contrast0}) (solid curve), along with the corresponding upper and lower $95\%$ credible interval (CrI) (dashed curves), 
 as a function of the posterior mean of $\bm{x}^\top\bm{\beta}$. 
 The right panel  also displays the expected odds ratio (CCP vs. control) (solid curve) and the corresponding $95\%$ CrI (dashed curves), 
 but the horizontal axis is now the treatment-benefit-index (TBI) (\ref{TBI}), $P(\exp(\Delta(\bm{x})) < 1)$, 
 where $\exp(\Delta(\bm{x}))$ represents the odds ratio, 
 and the TBI probability is evaluated with respect to the posterior distribution of the parameters in $\Delta$. 
 The TBI provides a gradient of benefit that ranges from $0$ to $1$, with a higher value of the TBI indicating a greater benefit from the CCP treatment, compared to control.  
 The observed values for the quantities on the horizontal axes 
 are represented by the small blue ticks. 
} \label{fig1}
\end{figure}

  In the first panel of Figure~\ref{fig1}, 
  we display the individualized treatment effect, $\Delta(\bm{x}, \bm{\theta})$, 
 as a function of the single-index $\bm{x}^\top \bm{\beta}$.  
Specifically, we display the posterior mean of $\bm{x}^\top \bm{\beta}$ and the values $\bm{x}_i^\top \bm{\beta}$ $(i=1,\ldots,n)$ $(n=2287)$ on the horizontal axis, 
  where these ``observed'' values  are represented by the small blue ticks on the horizontal axis. 
 The uncertainty in the estimation of the single-index coefficient $\bm{\beta}$ (as well as that of $\bm{m}$) is also accounted for in the credible bands in Figure~\ref{fig1}. 
 For the interpretability, we exponentiate the HTE estimate $\Delta(\bm{x}, \bm{\theta}) = g(\bm{x}^\top\bm{\beta}, A=1) - g(\bm{x}^\top\bm{\beta}, A=0)$, 
 so that the vertical axis in the panel represents the odds ratio (CCP vs. control) for a bad outcome (mechanical ventilation or death). 
An odds ratio of less than $1$ indicates a superior CCP efficacy over the control treatment.  
 As most of the observed values $\bm{x}_i^\top \bm{\beta}$ of the single-index 
 fall below the line representing the odds ratio of $1$, 
 most of the patients are expected to benefit from CCP treatment, 
 except those with the $\bm{x}_i^\top \bm{\beta}$ values greater than  $0.45$, 
 where their corresponding expected individualized odds ratios are greater than $1$ (about $28\%$ of the observed patients). 
 The $U$-shaped nonlinear relationship between the odds ratio and the single-index of the model suggests that 
  the use of the flexible link function $g$ in (\ref{eq2}) is more adequate than using a more restricted linear model for this HTE modeling.

\begin{table}[H]
	\caption{Pretreatment patient characteristics $\bm{X}$ and the corresponding estimated index coefficients $\bm{\beta}$ (and $95\%$ CrI)}
	\centering
	\begin{tabular}{lcc}
		\toprule
		Pretreatment characteristic $x_j$    & Index coefficient $\beta_j$ \ [$95\%$ CrI]  \\ 
		\midrule
		Oxygen by mask or nasal prongs$^\ast$ (1/0) & 0.68 \ [0.50, 0.80]   \\
		Oxygen by high flow$^\ast$ (1/0)   & 0.47 \ [0.16, 0.61]      \\
		Age (dichotomized, $\ge 67$) (1/0)     & -0.13 \ [-0.46,0.04]       \\
		Blood type (A or AB vs. O or B) (1/0)     & -0.31 \ [-0.49, -0.16]      \\
		Cardiovascular disease  (1/0)   & -0.24  \  [-0.65,-0.06]    \\
		Diabetes  (1/0)   & -0.26  \ [-0.52, -0.08]     \\
		Pulmonary disease  (1/0)   & 0.05  \ [-0.16,0.22]     \\
		\bottomrule
		$^\ast$ The reference level: hospitalized but no oxygen therapy required. 
	\end{tabular}
	\label{tab:table}
\end{table}

Although the first panel of Figure~\ref{fig1} displays 
a useful information about the relationship between the individualized treatment effect
$\exp(\Delta(\bm{x}))$ (i.e., the  individualized  odds ratio) and  the posterior mean of the single-index $\bm{x}^\top\bm{\beta}$, 
this relationship is non-monotonic, which makes it difficult to construct a ``gradient'' of the treatment benefit from $A=1$ vs $A=0$, 
as a function of the patient characteristics $\bm{x}$. 
Thus, in the second panel of Figure~\ref{fig1}, 
we display the individualized odds ratio $\exp(\Delta(\bm{x}))$, 
as a function of the TBI defined in (\ref{TBI}), i.e., $P(\exp(\Delta(\bm{x})) < 1)$, 
where  the probability is evaluated with respect to the posterior distribution of the parameters involving $\Delta$. 
As a probability, the TBI ranges from $0$ to $1$: larger values are associated with larger CCP benefit. 
For example, for the patients with a large value of the TBI (i.e., 
 TBI scores near $1$) were expected to experience large, clinically meaningful benefits from CCP.

The second panel of Figure~\ref{fig1} displays a monotonically decreasing trend of the expected odds ratio (an increasing CCP benefit), as the TBI score increases from $0$ to $1$.  
 Some portions of the expected odds ratio and the corresponding $95\%$ CrI exceed $1$ for very small TBI values, suggesting the possibility of harm from CCP as the TBI approaches $0$, 
 whereas the TBI values close to $1$ indicate a substantial benefit from the CCP treatment over the control treatment.  
We can use the TBI score to stratify patients according to their predicted treatment benefit levels.



\section{Discussion} 

The idea in the Bayesian estimation approach of \cite{Antoniadis} was to treat the link function $g$ as another unknown and approximate it by a linear combination of $B$-spline basis functions. In this article, to model heterogeneous treatment effect using a flexible link function,  in (\ref{priorbeta2}),  we specify the prior for the $B$-spline coefficient $\bm{\gamma}$, conditional on $\bm{\beta}$, as normal with the same dispersion matrix as the WLS  estimator (i.e., a Zellner's g-prior) defined based on the adjusted responses and the weights associated with the first step of IWLS, for each sampler.  The approximation under the IWLS framework and the specific prior choice (\ref{priorbeta2}) allows us to analytically integrate $\bm{\gamma}$ out of the approximated posterior (\ref{tmp1}), which simplifies the sampling procedure for $\bm{\beta}$. Although the sampling was done using  approximated conditional posteriors, this approach appeared to work reasonably well. 

\bibliographystyle{unsrtnat} 
\bibliography{refs}  







\end{document}